\documentclass[twocolumn,showpacs,preprintnumbers,amsmath,amssymb]{revtex4}
\usepackage{graphicx}% Include figure files
\usepackage{dcolumn}% Align table columns on decimal point
\usepackage{bm}% bold math

\usepackage{epsfig}
\usepackage{rotating}

\begin{document}
%\preprint{APS/123-QED}
\title{\boldmath Partial wave analysis of $J/\psi \rightarrow p \bar p
\pi^0$}% Force line breaks with \\

\author
{M.~Ablikim$^{1}$,              J.~Z.~Bai$^{1}$,   Y.~Bai$^{1}$,
Y.~Ban$^{11}$, X.~Cai$^{1}$,                  H.~F.~Chen$^{16}$,
H.~S.~Chen$^{1}$,              H.~X.~Chen$^{1}$, J.~C.~Chen$^{1}$,
Jin~Chen$^{1}$
% ...
\\
\vspace{0.2cm}
(BES Collaboration)\\
\vspace{0.2cm}
{\it
$^{1}$ Institute of High Energy Physics, Beijing 100049, People's Republic of China\\
% ...
$^{11}$ Peking University, Beijing 100871, People's Republic of China\\
% ...
$^{16}$ University of Science and Technology of China, Hefei 230026,
People's Republic of China\\
% ...
}
}

\date{\today}% It is always \today, today,
             %  but any date may be explicitly specified

\begin{abstract}
Using a sample of 58 million $J/\psi$ events collected with the BESII
detector at the BEPC, more than 100,000 $J/\psi \rightarrow p\bar p
\pi^0$ events are selected, and a detailed partial wave analysis is
performed. The branching fraction is determined to be $Br(J/\psi
\rightarrow p \bar p \pi^0)=(1.33 \pm 0.02 \pm 0.11) \times
10^{-3}$. A long-sought `missing' $N^*$, first observed in $J/\psi
\rightarrow p \bar n \pi^-$, is observed in this decay too, with
mass and width of $2040_{-4}^{+3}\pm 25$ MeV/c$^2$ and
$230_{-8}^{+8}\pm 52$ MeV/c$^2$, respectively.  Its spin-parity favors
$\frac{3}{2}^+$. The masses, widths, and spin-parities of other $N^*$
states are obtained as well.

\end{abstract}
\pacs{13.25.Gv, 12.38.Qk, 14.20.Gk, 14.40.Cs}
%\pacs{13.20.Gd, 12.38.Qk, 14.40.Cs}% PACS, the Physics and Astronomy
                             % Classification Scheme.
                 %\keywords{Suggested keywords}%Use showkeys class option if keyword
                              %display desired
\maketitle

\section{Introduction}

Studies of mesons and searches for glueballs, hybrids, and multiquark
states have been active fields of research since the early days of
elementary particle physics. However, our knowledge of baryon
spectroscopy has been poor due to the complexity of the three quark
system and the large number of states expected.

As pointed out by N. Isgur \cite{isgur} in 2000, nucleons are the
basic building blocks of our world and the simplest system in which
the three colors of QCD neutralize into colorless objects and the
essential non-abelian character of QCD is manifest, while baryons are
sufficiently complex to reveal physics hidden from us in the mesons.
The understanding of the internal quark-gluon structure of baryons is
one of the most important tasks in both particle and nuclear physics, and
the systematic study of baryon spectroscopy, including production and
decay rates, will provide important information in understanding the
nature of QCD in the confinement domain.

In recent years, interest in baryon spectroscopy has revived. For
heavy baryons containing a charm or bottom quark, new exciting results
have been obtained since the experimental evidence for the first
charmed baryon $\Sigma_c^{++}$ was reported by BNL \cite{bnl} in 1975
in the reaction $\nu_{\mu} p \to \mu^- \Lambda \pi^+ \pi^- \pi^+
\pi^-$. Many charmed baryons have been observed in recent years in
CLEO, the two B-factories, the Fermilab photo-production experiment,
FOCUS, and SELEX \cite{cb1,cb2,cb3,cb4,cb5}. Only a few baryons with
beauty have been discovered so far. Earlier results on beauty baryons
were from CERN ISR and LEP \cite{bb1} experiments, while new beauty
baryons are from CDF and D0 at the Tevatron \cite{bb2,bb3,cb5}. Most
information on light-quark baryons comes from $\pi N$ or $K N$ elastic
or charge exchange scattering, but new results are being added from
photo- and electro-production experiments at JLab and the ELSA, GRAAL,
SPRING8, and MAMI experiments, as well as $J/\psi$ and $\psi(2S)$
decays at BES. However, up to now, the available experimental
information is still inadequate and our knowledge on $N^*$ resonances
is poor. Even for the well-established lowest excited states,
$N(1440)$, $N(1535)$, {\it etc.}, their properties, such as masses,
widths, decay branching fractions, and spin-parity assignments, still
have large experimental uncertainties \cite{pdg}. Another outstanding
problem is that, the quark model predicts a substantial number of
$N^*$ states around 2.0 GeV/c$^2$ \cite{scaw,nig,scaw2}, but some of
these, the `missing' $N^*$ states, have not been observed
experimentally.

$J/\psi$ decays provide a good laboratory for studying not only
excited baryon states, but also excited hyperons, such as $\Lambda^*$,
$\Sigma^*$, and $\Xi^*$ states. All $N^*$ decay channels which are
presently under investigation in photo- and electro-production
experiments can also be studied in $J/\psi$ decays. Furthermore, for
$J/\psi \to N \bar N \pi$ and $N \bar N \pi \pi$ decays, the
$N\pi(\bar{N}\pi)$ and $N\pi\pi(\bar N\pi\pi)$ systems are expected to
be dominantly isospin 1/2 due to that the isospin conserving
three-gluon annihilation of the constituent c-quarks dominates over
the isospin violating decays via intermediate photon for the baronic
final states. This makes the study of $N^*$ resonances from $J/\psi$
decays less complicated, compared with $\pi N$ and $\gamma N$
experiments which have states that are a mixture of isospin 1/2 and
3/2.

$N^*$ production in $J/\psi \to p \bar p \eta$ was studied using a
partial wave analysis (PWA) with $7.8 \times 10^6 J/\psi$ BESI
events~\cite{plb75}. Two $N^*$ resonances were observed with
masses and widths of $M=1530 \pm 10$ MeV, $\Gamma=95 \pm 25$ MeV and
$M=1647 \pm 20 $MeV, $\Gamma=145 ^{+80}_{-45} $ MeV, and spin-parities
favoring $J^{P} = \frac{1}{2}^-$. In a recent analysis of $J/\psi \rightarrow
p \bar n \pi^- + c.c.$ \cite{xbji}, a `missing' $N^*$ at around 2.0
GeV/c$^2$ named $N_{x}(2065)$ was observed, based on $5.8 \times 10^7
J/\psi$ events collected with BESII at the Beijing
Electron Positron Collider (BEPC). The mass and width for this state
are determined to be $2065\pm 3 _{-30}^{+15}$ MeV/c$^2$ and $175\pm
12\pm 40$ MeV/c$^2$, respectively, from a simple Breit-Wigner fit. In
this paper, the results of a partial wave analysis of $J/\psi
\rightarrow p \bar p \pi^0$ are presented, based on the same event sample.

\section{Detector and data samples}

The upgraded Beijing Spectrometer detector, is a
large solid-angle magnetic spectrometer which is described in
detail in Ref. \cite{bes2}. The momenta of charged particles
are determined by a 40-layer cylindrical main drift chamber(MDC)
which has a momentum resolution of $\sigma_{p}/p=1.78\%
\sqrt{1+p^2}$ (p in GeV/c). Particle identification is
accomplished by specific ionization ($dE/dx$) measurements in the
drift chamber and time-of-flight (TOF) information in a
barrel-like array of 48 scintillation counters. The $dE/dx$
resolution is $\sigma_{dE/dx}=8.0\%$; the TOF resolution for
Bhabha events is $\sigma_{TOF}=180$ ps. A 12-radiation-length
barrel shower counter (BSC) comprised of gas tubes interleaved
with lead sheets is radially outside of the time-of-flight
counters. The BSC measures the energy and direction of photons
with resolutions of $\sigma_{E}/E \simeq 21\%/\sqrt{E}$ ($E$ in
GeV), $\sigma_{\phi}=7.9$ mrad, and $\sigma_{z}=2.3$ cm. Outside
of the solenoidal coil, which provides a 0.4~Tesla magnetic field
over the tracking volume, is an iron flux return that is
instrumented with three double layers of counters that identify
muons of momenta greater than 0.5~GeV/c.

In this analysis, a GEANT3-based Monte Carlo (MC) program, with
detailed consideration of detector performance is used. The
consistency between data and MC has been carefully checked in many
high-purity physics channels, and the agreement is reasonable. More
details on this comparison can be found in Ref. \cite{simbes}.
%The signal channel $\jpsito \ppb \piz$, $\piz \to 2\gamma$ is
%generated with a phase space generator. For this decay channel,
%1.16M events are simulated. To study possible background in our
%analysis,100K $\gamma p \bar p$, 140K $p \bar p \pi^0 \pi^0$, 40K $p
%\bar p \omega,\omega \rightarrow \gamma \pi^0$, 100K $p \bar p \eta
%$ and 100K $p \bar p  $ events are generated.

\section{Event selection}

\begin{figure*}
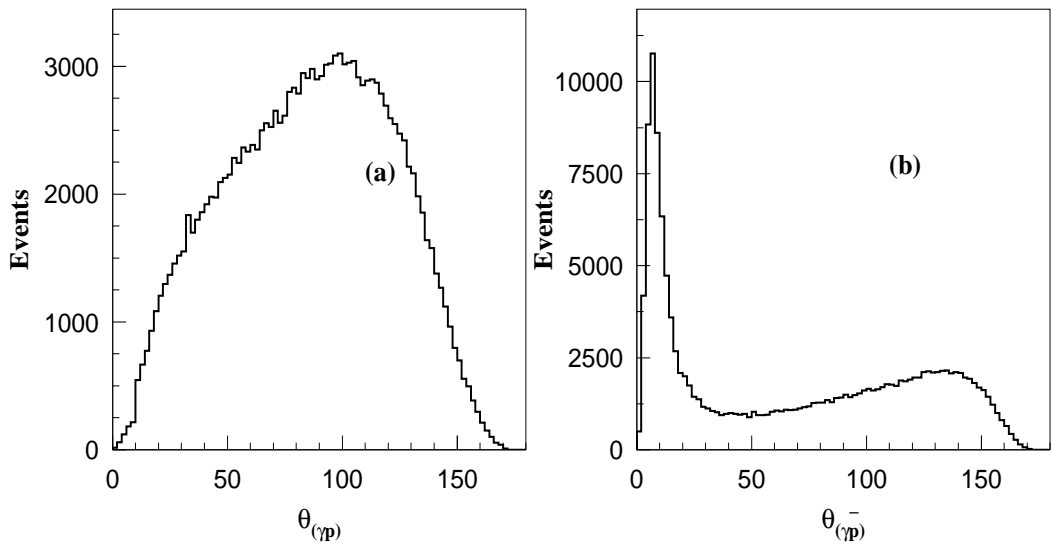

\includegraphics[height=2.8in,width=2.7in]{figu/ang1a.epsi}
\includegraphics[height=2.8in,width=2.7in]{figu/ang1b.epsi}
\caption{\label{fig:bepc}Distributions of (a) $\theta_{\gamma p}$
  and (b) $\theta_{\gamma \bar p}$ in $J/\psi \rightarrow p \bar p
  \pi^0$ MC simulation. $\theta_{\gamma p}$ and $\theta_{\gamma \bar
    p}$ are the angles between the $p$ or $\bar p$ and the nearest
  neutral cluster. }
\end{figure*}

The decay $J/\psi \rightarrow p \bar p \pi^0$ with $\pi^0 \to
\gamma \gamma$ contains two charged tracks and two photons. The
first level of event selection for $J/\psi\to p \bar p \pi^0$
candidate events requires two charged tracks with total charge
zero.  Each charged track, reconstructed using MDC information, is
required to be well fitted to a three-dimensional helix, be in the
polar angle region $|\cos\theta_{MDC}|<0.8$, and have the point of
closest approach of the track to the beam axis to be within 1.5 cm
radially and within 15 cm from the center of the interaction
region along the beam line. More than two photons per candidate
event are allowed because of the possibility of fake photons
coming from interactions of the charged tracks in the detector,
from $\bar p$ annihilation, or from electronic noise in the shower
counter. A neutral cluster is considered to be a photon candidate
when the energy deposited in the BSC is greater than 50~MeV, the
angle between the nearest charged tracks and the cluster is
greater than 10$^{\circ}$, and the angle between the cluster
development direction in the BSC and the photon emission direction
is less than 23$^{\circ}$. Because of the large number of fake
photons from $\bar p$ annihilation, we further require the angle
between the $\bar p$ and the nearest neutral cluster be greater
than $20^{\circ}$. Figures~\ref{fig:bepc}~(a) and (b) show the
distributions of the angles $\theta_{\gamma p}$ and
$\theta_{\gamma \bar p}$ between the $p$ or $\bar p$ and the
nearest neutral cluster for $J/\psi \rightarrow p \bar p \pi^0$ MC
simulation; most of the fake photons from $\bar p$ annihilation
accumulate at small angles.

To identify the proton and antiproton, the combined TOF and
$dE/dx$ information is used. For each charged track in an event, the
particle identification (PID)
$\chi^{2}_{PID}(i)$ is determined using:
\begin{eqnarray}
\chi_{TOF}(i)=\frac{TOF_{measured}-TOF_{expected}(i)}{\sigma_{TOF}(i)}
\nonumber \\
\chi_{dE/dx}(i)=\frac{dE/dx_{measured}-dE/dx_{expected}(i)}{\sigma_{dE/dx}(i)}
\nonumber \\
\chi^{2}_{PID}(i) = \chi^{2}_{dE/dx}(i) + \chi^{2}_{TOF}(i), \nonumber
\end{eqnarray}
where $i$ corresponds to the particle hypothesis.  A charged track is
identified as a proton if $\chi^{2}_{PID}$ for the proton hypothesis
is less than those for the $\pi$ or $K$ hypotheses.  For the channel
studied, one charged track must be identified as a proton and the
other as an antiproton. The selected events are subjected to a 4-C
kinematic fit under the $J/\psi \rightarrow p \bar p \gamma \gamma$
hypothesis. When there are more than two photons in a candidate event,
all combinations are tried, and the combination with the smallest 4-C
fit $\chi^{2}$ is retained.

\begin{figure*}
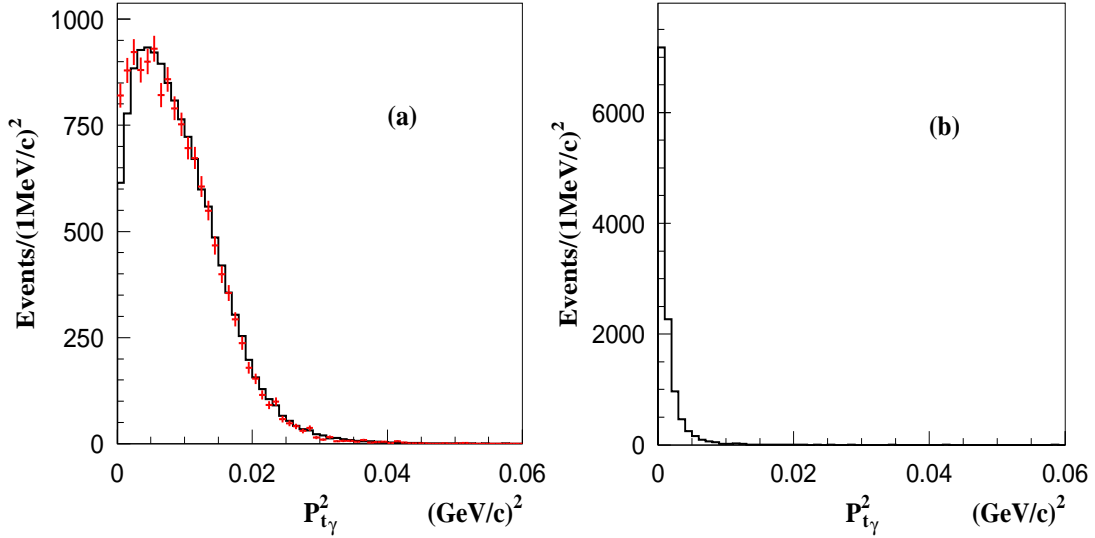

\includegraphics[height=2.8in,width=2.8in]{figu/ptr2a.epsi}
\includegraphics[height=2.8in,width=2.8in]{figu/ptr2b.epsi}
\caption{$P_{t\gamma}^2$ distributions. For (a), crosses
  are data, and the histogram is MC simulation of $J/\psi \rightarrow
  p \bar p \pi^0$. (b) Distribution for simulated $J/\psi \rightarrow
  \gamma p \bar p$ events.
  $P^2_{t\gamma}=4|\vec{P}_{miss}|^2\sin^2\theta_{\gamma}/2$ where
  $\vec{P}_{miss}$ is the missing momentum in the event determined
  using the two charged particles, and $\theta_{\gamma}$ is the angle
  between $\vec{P}_{miss}$ and the higher energy photon.}
\label{fig:ptr}
\end{figure*}

In order to reduce contamination from back-to-back decays, such as
$J/\psi \rightarrow p \bar p$ {\it etc.}, the angle between two
charged tracks, $\theta_{2chrg}$, is required to be less than
175$^{\circ}$.  Figures~\ref{fig:ptr}~(a) and (b) show the
distributions of $P^2_{t\gamma}$ for simulated $J/\psi \rightarrow
p \bar p \pi^0$ and $J/\psi \rightarrow \gamma p \bar p$ events,
respectively. Selected data events are shown in
Figure.~\ref{fig:ptr}~(a). Here, the variable $P^2_{t\gamma}$ is
defined as:
$P^2_{t\gamma}=4|\vec{P}_{miss}|^2\sin^2\theta_{\gamma}/2$ where
$\vec{P}_{miss}$ is the missing momentum in the event determined
using the two charged particles, and $\theta_{\gamma}$ the angle
between $\vec{P}_{miss}$ and the higher energy photon. By
requiring $P_{t\gamma}^2>$ 0.003 GeV$^2$/c$^2$, background from
$J/\psi \rightarrow \gamma p \bar p$ is effectively reduced.

\begin{figure}
\includegraphics[height=2.8in,width=2.8in]{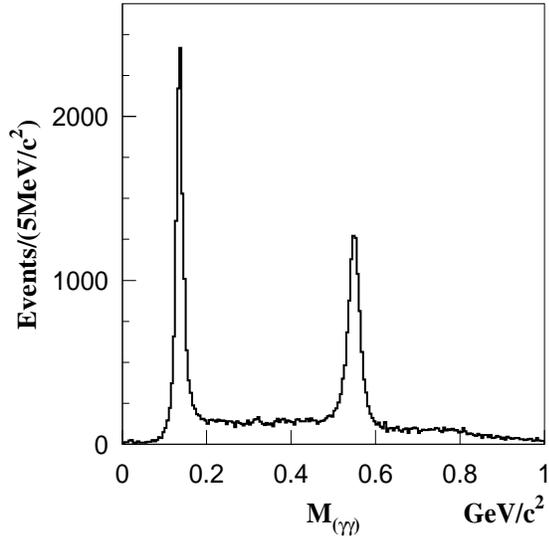}
\caption{The $\gamma \gamma$ invariant mass spectrum of $J/\psi
\rightarrow p \bar p \gamma \gamma$ candidates.}
\label{fig:2gam}
\end{figure}

\begin{figure*}
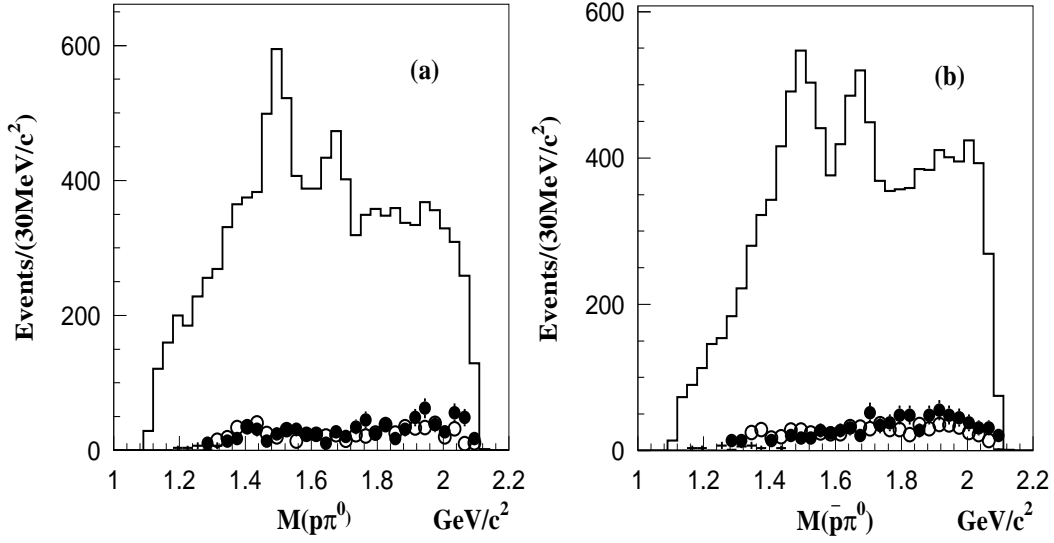

\includegraphics[height=2.8in,width=2.7in]{figu/ppb2pi0ppi4a.epsi}
\includegraphics[height=2.8in,width=2.7in]{figu/ppb2pi0ppi4b.epsi}
\caption{The invariant mass spectra of (a) $M_{p\pi^0}$ and (b) $M_{\bar
p \pi^0}$ for $J/\psi \rightarrow p \bar p \pi^0$ candidate events, where the
circles with error bars are the background events estimated from
$\pi^0$ sideband events, and the black dots with error bars are those
from simulated $J/\psi \rightarrow p \bar p \pi^0 \pi^0$ events
passing the selection criteria. } \label{fig:ppi}
\end{figure*}

\begin{figure}
\includegraphics[height=2.8in,width=2.8in]{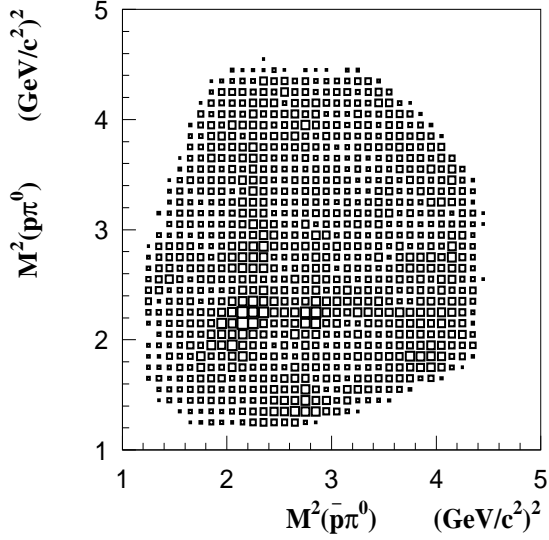}
\caption{Dalitz plot of $J/\psi \rightarrow p \bar p \pi^0$ candidates.}
\label{fig:dalitz}
\end{figure}

\begin{figure}
\includegraphics[height=2.8in,width=2.8in]{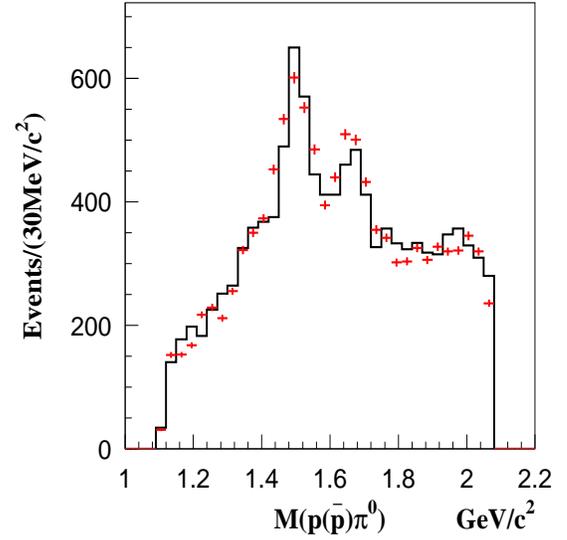}
\caption{The re-normalized invariant mass spectra of $M_{p\pi^0}$ and $M_{\bar p
\pi^0}$ after correction for detection efficiency, where the
histogram is $M_{p \pi^0}$ and the crosses are $M_{\bar p \pi^0}$.}
\label{fig:eff1}
\end{figure}

The $\gamma \gamma$ invariant mass spectrum after the above selection
criteria is shown in Fig.~\ref{fig:2gam}, where $\pi^0$ and $\eta$
signals can be seen clearly. To select $J/\psi \to p \bar p \pi^0$
events, $|M_{\gamma \gamma}-0.135|<0.03$ GeV$/c^2$ is
required. Figures~\ref{fig:ppi}~(a) and (b) show the invariant mass
spectra of $M_{p\pi^0}$ and $M_{\bar p \pi^0}$, respectively, and
clear $N^*$ peaks are seen at around 1.5 GeV/c$^2$ and 1.7
GeV/c$^2$. The Dalitz plot of this decay is shown in
Fig.~\ref{fig:dalitz}, and some $N^*$ bands are also evident. Both the
mass spectra and Dalitz plot exhibit an asymmetry for $m_{p \pi^0}$
and $m_{\bar p \pi^0}$, which is mainly caused by different detection
efficiencies for the proton and antiproton. The re-normalized $M_{p \pi^0}$ and
$M_{\bar p \pi^0}$ invariant mass spectra after efficiency
corrections are shown as the solid histogram and crosses,
respectively, in Fig. \ref{fig:eff1}, and the agreement is better.
%, but the discrepancy exists because of
%a not-so-good MC simulation for $\bar p$ .

Other possible $J/\psi \to p \bar p \pi^0$ backgrounds are studied
using MC simulation and data. Decay channels that have similar final
states as $J/\psi \to p \bar p \pi^0$ are simulated, and $J/\psi \to p
\bar p \pi^0 \pi^0$ is found to be the main background channel.
Surviving $J/\psi \rightarrow p \bar p \pi^0 \pi^0$ events, passing
all requirements described above, are plotted as black dots in
Fig.~\ref{fig:ppi}. The invariant mass distribution of this background
can be described approximately by phase space. The $\pi^0$ sideband,
defined by  $0.2 < (M_{\gamma \gamma}-0.135) <0.2278$ GeV$/c^2$, is
used to estimate the background from non-$\pi^0$ final states, such as
$J/\psi \rightarrow \gamma p \bar p$, {\it etc.}. The circles in
Fig.~\ref{fig:ppi}~ show the contribution from $\pi^0$ sideband
events. In the partial wave analysis, described below, two kinds of
background are considered, $\pi^0$ sideband background and a
non-interfering phase space background to account for the background
from $J/\psi \rightarrow p \bar p \pi^0 \pi^0$.

\section{Partial wave analysis}
A partial wave analysis (PWA) is performed to study the $N^*$ states
in this decay. The sequential decay process can be described by
$J/\psi \to \bar p N^*(p \bar N^*)$, $N^*(\bar N^*) \to p \pi^0 (\bar p
\pi^0)$. The amplitudes are constructed using the relativistic
covariant tensor amplitude formalism \cite{wrjs,whl}, and the maximum
likelihood method is used in the fit.
\subsection{Introduction to PWA}
\ The basic procedure for the partial wave analysis is the
standard maximum likelihood method:\\
\ (1) Construct the amplitude $A_{j}$ for the $j$-th possible partial
wave in $J/\psi \rightarrow p \bar N_{X},\bar N_{X} \rightarrow
\bar p \pi^0$ or $J/\psi \rightarrow \bar p N_{X},N_{X} \rightarrow
p \pi^0$ as:
\begin{eqnarray}
A_{j}=A_{prod-X}^{j} (BW)_{X}A_{decay-X},
\end{eqnarray}
where $A_{prod-X}^{j}$ is the amplitude which describes the production of
the intermediate resonance $N_{X}$, $BW_{X}$ is the Breit-Wigner propagator of
$N_{X}$, and $A_{decay-X}$ is the decay amplitude of $N_{X}$.
The corresponding term
for the $\bar N_{X}$ is obtained by charge conjugation
with a negative sign due to negative C-parity of $J/\psi$. \\
\ (2) The total transition probability, $\omega$, for each event is
obtained from the linear combination of these partial wave amplitudes
$A_{j}$ as $\omega=|\Sigma_{j} c_{j}A_{j}|^2$, where the $c_{j}$
parameters are to
be determined by fitting the data.\\
\ (3) The differential cross section is given by:
\begin{eqnarray}
\frac{d \sigma}{d \Phi}=|\Sigma_{j} c_{j}A_{j}|^2+F_{bg},
\end{eqnarray}
where, $F_{bg}$ is the background function, which includes
$\pi^0$ sideband background and non-interfering phase space background. \\
\ (4) Maximize the following likelihood function $ln\cal L$ to
obtain $c_{j}$ parameters, as well as the masses and widths of the
resonances.
\begin{eqnarray}
{ln \cal L}= \sum \limits_{k=1}^{n} ln \frac{ \omega(\xi_{k})}
{\int d \xi \omega(\xi)\epsilon(\xi)},
\end{eqnarray}
where $\xi_{k}$ is the energy-momentum of the final state of the $k$-th
observed event, $\omega(\xi)$ is the probability to generate the
combination $\xi$, $\epsilon(\xi)$ is the detection efficiency for
the combination $\xi$. As is usually done,
rather than maximizing $\mathcal{L}$, $\mathcal{S} = -\rm{ln}
\mathcal{L}$ is minimized.

\ For the construction of partial wave amplitudes, we assume the
effective Lagrangian approach~\cite{mbnc,mgoet} with the
Rarita-Schwinger formalism~\cite{wrjs,cfncs,suc3,suc2}. In this
approach, there are three basic elements for constructing
amplitudes: the spin wave functions for particles, the propagators,
and the effective vertex couplings. The amplitude can then be written
out by Feynman rules for tree diagrams.

\ For example, for $J/\psi \rightarrow \bar N
N^*(\frac{3}{2}^+) \rightarrow \bar N(\kappa_{1},s_{1})N(\kappa_{2},s_{2})
\pi(\kappa_{3})$, the amplitude can be constructed as:
%\begin{widetext}
\begin{eqnarray}
A_{\frac{3}{2}^+}=&&\bar
u(\kappa_{2},s_{2})\kappa_{2\mu}P_{3/2}^{\mu\nu}(c_{1}g_{\nu\lambda}+c_{2}\kappa
_{1\nu}\gamma_{\lambda}\nonumber\\
&&+c_{3}\kappa_{1\nu}\kappa_{1\lambda})\gamma_{5}\upsilon
(\kappa_{1},s_{1})\psi^{\lambda},
\end{eqnarray}
%\end{widetext}
where $u(\kappa_{2},s_{2})$ and $\upsilon (\kappa_{1},s_{1})$ are
$\frac{1}{2}$-spinor wave functions for $N$ and $\bar N$,
respectively; $\psi^{\lambda}$ is the spin-1 wave function, $i.e.$,
the polarization vector for $J/\psi$. The $c_{1}$, $c_{2}$, and
$c_{3}$ terms correspond to three possible couplings for the $J/\psi
\rightarrow \bar N N^*(\frac{3}{2}^+)$ vertex. They can be taken as
constant parameters or as smoothly varying vertex form factors. The
spin $\frac{3}{2}^+$ propagator $P_{3/2+}^{\mu\nu}$ for
$N^*(\frac{3}{2}^+)$ is:

\begin{eqnarray}
P_{3/2+}^{\mu\nu}=&&\frac{\gamma\cdot
p+M_{N^*}}{M_{N^*}^2-p^2+iM_{N^*}\Gamma_{N^*}}[g^{\mu\nu}-\frac{1}{3}\gamma^{\mu}
\gamma^{\nu}\nonumber\\
&&-\frac{2p^{\mu}p^{\nu}}{3M_{N^*}^2}+\frac{p^{\mu} \gamma^{\nu}-
p^{\nu} \gamma^{\mu}}{3M_{N^*}}],
\end{eqnarray}
with $p=\kappa_{2}+\kappa_{3}$. Other partial wave amplitudes can be constructed
similarly~\cite{wrjs,whl}.

%\subsection{Intermediate resonances considered }

The possible intermediate resonances are listed in
Table~\ref{tab:res}. Of these states, only a few are (well)
established states, while $N_x(1885)$ is one of the `missing' $N^*$
states predicted by the quark model and not yet experimentally
observed. $N_x(2065)$ is also a long-sought `missing' $N^*$,
which was observed recently by BES \cite{xbji}.

%\subsection{Breit-Wigner formulae }

For the lowest lying $N^*$ states, $N(1440)$, $N(1520)$, and
$N(1535)$, Breit-Wigner's with phase space dependent widths
are used.

\begin{equation}
BW_{X}(s)=\frac{m \Gamma(s)}{s-m^2+im \Gamma(s)},
\end{equation}
where $s$ is the invariant mass-squared. The phase space dependent widths
can be written as \cite{tpv181}:

\begin{eqnarray}
\Gamma_{N(1440)(s)} =&& \Gamma_{N(1440)}(0.7 \frac{B_{1}(q_{\pi N})
\rho_{\pi N}(s)}{B_{1}(q_{\pi N}^{N^*}) \rho_{\pi
N}(M_{N^*}^2)}\nonumber\\
&&+0.3\frac{B_{1}(q_{\pi \Delta}) \rho_{\pi \Delta}(s)}{B_{1}(q_{\pi
\Delta}^{N^*}) \rho_{\pi \Delta}(M_{N^*}^2)}),
\end{eqnarray}

\begin{eqnarray}
\Gamma_{N(1520)} =&& \Gamma_{N(1520)} \frac{B_{2}(q_{\pi N})
\rho_{\pi N}(s)}{B_{2}(q_{\pi N}^{N^*}) \rho_{\pi N}(M_{N^*}^2)},
\end{eqnarray}

\begin{eqnarray}
\Gamma_{N(1535)} =&&\Gamma_{N(1535)}(0.5 \frac{\rho_{\pi
N}(s)}{\rho_{\pi N}(M_{N^*}^2)}\nonumber\\
&&+0.5 \frac{\rho_{\eta N}(s)}{\rho_{\eta N}(M_{N^*}^2)}),
\end{eqnarray}
 where $B_{l}(q)$ ($l=1,2$) is
the standard Blatt-Weisskopf barrier factor \cite{suc3,suc2} for the
decay with orbital angular momentum $L$ and $\rho_{\pi N}(s)$,
$\rho_{\pi \Delta}(s)$, and $\rho_{\eta N}(s)$ are the phase space
factors for $\pi N$, $\pi \Delta$, and $\eta N$ final states,
respectively.
\begin{equation}
\rho_{XY}(s)=\frac{2q_{XY}(s)}{\sqrt{s}},
\end{equation}

\begin{eqnarray}
q_{XY}(s)= \frac{\sqrt{(s-(M_{Y}+M_{X})^2)(s-(M_{Y}-M_{X})^2)}}
{(2\sqrt{s})},
\end{eqnarray}
where $X$ is $\pi$ or $\eta$, $Y$ is $N$ or $\Delta$, and
$q_{XY}(s)$ is the momentum of $X$ or $Y$ in the center-of-mass
(CMS) system of $XY$. For other resonances, constant width
Breit-Wigner's are used.

As described in Ref.~\cite{liangwh}, the form factors are
introduced to take into account the nuclear structure. We have tried
different form factors, given in Ref. \cite{liangwh}, in the
analysis and find that for $J= \frac{1}{2}$ resonances, the form
factor preferred in fitting is
\begin{equation}
F_{N}(s_{\pi N})= \frac{\Lambda_{N}^4}{\Lambda_{N}^4+(s_{\pi
N}-m_{N^*}^2)^2},
\end{equation}
where $s_{\pi N}$ is the invariant mass squared of $N$, $\pi$,
and for $J= \frac{3}{2}$ or $\frac{5}{2}$ states, the preferred form
factor is

\begin{equation}
F_{N}(s_{\pi N})=e^\frac{-|s_{N \pi}-m_{N^*}^2|}{\Lambda^2}.
\end{equation}
Therefore, the above form factors are used in this analysis.

%For the background in this decay, two kinds of background are
%onsidered in PWA analysis, as is mentioned above.
In the log likelihood calculation, $\pi^0$ sideband background events
are given negative weights; the sideband events then cancel background
in the selected candidate sample. The $J/\psi \to p \bar p \pi^0
\pi^0$ background is described by a
non-interfering phase space term, and the amount of this background is
floated in the fit.

\begin{table}
\caption{\label{tab:res}Resonances considered in the PWA analysis.}
\begin{ruledtabular}
\begin{tabular}{ccccc}
Resonance & Mass(MeV) & Width(MeV) & $J^P$ & C.L.\\
\hline
$N(940)$& 940 & 0&$\frac{1}{2}^+$&off-shell\\
$N(1440)$& 1440 & 350&$\frac{1}{2}^+$&****\\
$N(1520)$& 1520 & 125&$\frac{3}{2}^-$&****\\
$N(1535)$& 1535 & 150&$\frac{1}{2}^-$&****\\
$N(1650)$& 1650 & 150&$\frac{1}{2}^-$&****\\
$N(1675)$& 1675 & 145&$\frac{5}{2}^-$&****\\
$N(1680)$& 1680 & 130&$\frac{5}{2}^+$&****\\
$N(1700)$& 1700 & 100&$\frac{3}{2}^-$&***\\
$N(1710)$& 1710 & 100&$\frac{1}{2}^+$&***\\
$N(1720)$& 1720 & 150&$\frac{3}{2}^+$&****\\
$N_{x}(1885)$& 1885 & 160&$\frac{3}{2}^-$&`missing' $N^*$\\
$N(1900)$& 1900 & 498&$\frac{3}{2}^+$&**\\
$N(2000)$& 2000 & 300&$\frac{5}{2}^+$&**\\
$N_{x}(2065)$& 2065 & 150&$\frac{3}{2}^+$&`missing' $N^*$\\
$N(2080)$& 2080 & 270&$\frac{3}{2}^-$&**\\
$N(2090)$& 2090 & 300&$\frac{1}{2}^-$&*\\
$N(2100)$& 2100 & 260&$\frac{1}{2}^+$&*\\
\end{tabular}
\end{ruledtabular}
\footnotetext{**** Existence is certain, and properties are at least
fairly well\\     $~~~~~~~~~~$    explored.}
\footnotetext{***$~$ Existence ranges from very likely to certain,
 but further  \\  $~~~~~~~~~~~$      confirmation is desirable and/or quantum numbers,
 branch-\\ $~~~~~~~~~~$ ing fractions, etc. are not well determined. }
\footnotetext{** $~~$ Evidence of existence is only fair.}
\footnotetext{* $~~~$ Evidence of existence is poor.}
\end{table}

\subsection{PWA results}

\begin{figure}
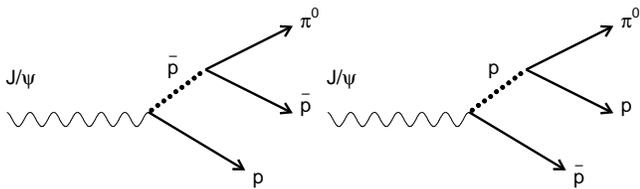

\includegraphics[height=0.95in]{figu/diag7a.epsi}
\includegraphics[height=0.95in]{figu/diag7b.epsi}
\caption{Feynman diagrams of $J/\psi \rightarrow p \bar p
\pi^0$ for the off-shell decay process.} \label{feynman}
\end{figure}

Well established states, such as $N(1440)$, $N(1520)$,
$N(1535)$, $N(1650)$, $N(1675)$, $N(1680)$ are included in this
partial wave
analysis. According to the framework of soft $\pi$ meson theory
\cite{larfd}, the off-shell decay process is also needed in this decay,
and therefore $N(940)$ ($M = 940$ MeV/c$^2$, $\Gamma$ = 0.0
MeV/c$^2$) is also included.  Fig.~\ref{feynman} shows the Feynman
diagram for this process.

\

\subsubsection{Resonances in the 1.7 GeV/c$^2$ mass region}

In the $M=1.7$ GeV/c$^2$ mass region, three resonances
$N(1700)$($\frac{3}{2}^-$), $N(1710)$($\frac{1}{2}^+$), and
$N(1720)$($\frac{3}{2}^+$) \cite{pdg} are supposed to decay into
$p\pi(\bar p \pi)$ final states. According to the Particle Data Group
(PDG08)~\cite{pdg}, only $N(1720)$ is a well established state. We now
study whether these three states are needed in $J/\psi \to p \bar p
\pi^0$. This is investigated for two cases, first assuming no $N^*$
states in the high mass region ($>$ 1.8 GeV/c$^2$), and second
assuming $N_{x}(2065)$, $N(2080)$, and $N(2100)$ states in the high
mass region. With no $N^*$ states in the $M >$ 1.8 GeV/c$^2$ mass
region, the PWA shows that the significances of $N(1700)$ and
$N(1720)$ are 3.2$\sigma$ ($\Delta S = 11$) and 0.8$\sigma$ ($\Delta S
= 3$), and their fractions are 0.3\% and 6\%, respectively; only
$N(1710)$ is significant. When $N_{x}(2065)$, $N(2080)$, and $N(2100)$
are included, the $N(1710)$ makes the log likelihood value $S$ better
by 65, which corresponds to a significance much larger than 5$\sigma$.
However, neither the $N(1700)$ nor the $N(1720)$ is significant. We
conclude that the $N(1710)$ should be included in the PWA.

\subsubsection{$N_{x}(2065)$}

The $N_{x}(2065)$, a long-sought `missing' $N^*$ predicted by the
quark model, was observed in $J/\psi \rightarrow p \bar n \pi^- +c.c.$
\cite{xbji} with a mass of $2065\pm 3 _{-30}^{+15}$ MeV/c$^2$ and a
width of $175\pm 12\pm 40$ MeV/c$^2$, determined from a simple
Breit-Wigner fit. We investigate the need for the $N_{x}(2065)$ in
$J/\psi \to p \bar p \pi^0$. Including the $N(1440)$, $N(1520)$,
$N(1535)$, $N(1650)$, $N(1675)$, $N(1680)$, $N(1710)$ and the
off-shell decay in the PWA fit, different $N_{x}(2065)$ spin-parities
($J^P$) and different combinations of high mass resonances are
tried. If there are no other resonances in the high mass region, the
log likelihood value improves by 288, which corresponds to a significance
of greater than $5\sigma$, when a
$\frac{3}{2}^{+}$$N_{x}(2065)$ is added. Thus, the $N_{x}(2065)$ is
definitely needed in this case, and its mass and width are optimized
to be $M=2057_{-6}^{+4}$ MeV/c$^2$ and $\Gamma=220_{-12}^{+11}$
MeV/c$^2$.

The significance and spin-parity of $N_{x}(2065)$ is further checked
under the following four hypotheses (A, B, C and D) for the high mass
resonances. Case A has $N(2080)$ and $N(2100)$ included, case B
$N(2080)$ and $N(2000)$, case C $N(2000)$, $N(2080)$, and $N(2100)$,
and case D $N(2080)$, $N(2090)$, and $N(2100)$. The changes of the log
likelihood values ($\Delta S$), the corresponding significances, and
the fractions of $N_{x}(2065)$ are listed in Table ~\ref{tab:case}
when a $\frac{3}{2}^{+}$ $N_{x}(2065)$ is added in the four cases.  The
log likelihood values become better by 58 to 126 when $N_{x}(2065)$ is
included. Therefore, $N_{x}(2065)$ is needed all cases. The
differences of log likelihood values for different $N_{x}(2065)$ $J^P$
assignments for the four combinations are listed in Table
~\ref{tab:sp1}.  The assignment of $J^P = \frac{3}{2}^{+}$ gives the
best log likelihood value except for the cases where there is large
interference. Spin-parity of $\frac{3}{2}^+$ is favored for
$N_{x}(2065)$.

\begin{table}[htpb]
\caption{\label{tab:case} Changes of log likelihood values
($\Delta S$), the corresponding significances, and the fractions of
$N_{x}(2065)$, when $N_{x}(2065)$ is added in the four cases.}
\begin{ruledtabular}
\begin{tabular}{cccc}
Case & $\Delta S$  & significance & fraction (\%)\\
\hline
A &   126 & $\gg$ 5$\sigma$ & 23\\
B &  158  & $\gg$ 5$\sigma$ & 24\\
C &  79 & $\gg$ 5$\sigma$ & 16\\
D &  58 & $\gg$ 5$\sigma$ & 22\\
\end{tabular}
\end{ruledtabular}
\end{table}

\begin{table}[htpb]
\caption{\label{tab:sp1}Comparison of log likelihood values for
different $J^P$ assignments for $N_{x}(2065)$.}
\begin{ruledtabular}
\begin{tabular}{ccccccc}
{$J^P$} & $\frac{1}{2}^+$ & $\frac{1}{2}^-$ &
$\frac{3}{2}^+$ &
$\frac{3}{2}^-$ & $\frac{5}{2}^+$ & $\frac{5}{2}^-$  \\
\hline
A & 85.8 & 49.3 & 0.0 & -32.2\footnotemark[1] & -36.9\footnotemark[2] & 34.1\\
B & 5.0 & 68.5 & 0.0 & 54.3 & -12.1\footnotemark[3] & 6.3\\
C & 98.1 & 39.8 & 0.0 & 85.6 & 76.1 & 14.4\\
D & 44.2 & 45.2 &  0.0 & 25.0 & 36.2 & 38.0\\
\end{tabular}
\end{ruledtabular}
\footnotetext[1]{$~$780\% interference between $Nx(2605)$ and $N(2080)$.}
\footnotetext[2]{ $~$529\% interference between $N(1680)$ and $N(2000)$.}
\footnotetext[3]{$~$860\% interference between $N(1680)$ and $N(2000)$.}
\end{table}

\subsubsection{Other resonances in high mass region}

%As was mentioned above, the high mass region is very complicated.
In addition to the observed resonances, $N(2000)$, $N(2080)$, $N(2090)$
and $N(2100)$, as well as the $N_x(2065)$, there is another possible
`missing' $N^*$ state, $N_{x}(1885)$, which is predicted by theory
but not yet observed.
%Now, we study these high mass
%resonances. \\

a) $N_{x}(1885)$   \\

%The $N_{x}(1885)$ is another `missing' $N^*$ state predicted by
%quark model.
In the $p (\bar p) \pi^0$ invariant mass spectrum, shown in
Fig. \ref{fig:ppi}, no obvious peak is seen near 1.89 GeV/c$^2$. We
study whether this state is needed in the partial wave analysis for the four
cases. The significances are $1.3\sigma$ ($\Delta S
= 3.0$), $3.2\sigma$ ($\Delta S =
8.8$), $3.4\sigma$ ($\Delta S = 9.7$), and greater than $5\sigma$
($\Delta S = 28.0$) in cases A, B, C, and D, respectively, when a
$N_{x}(1885)$ is included. Thus, the statistical
significance is larger than 5$\sigma$ only in case D. In our final
fit, $N_{x}(1885)$ is not included. However, the difference of
including and not including it will be taken as a
systematic error.\\

b) $N(2000)$, $N(2080)$, $N(2090)$, and $N(2100)$\\

We next study whether $N(2000)$, $N(2080)$, $N(2090)$ and
$N(2100)$ are all significant in the decay.
%%%%fah - messed up below?
First, we add $N(2000)$, $N(2080)$, $N(2090)$, and $N(2100)$ one
at a time with $N(940)$, $N(1440)$, $N(1520)$, $N(1535)$,
$N(1650)$, $N(1675)$, $N(1680)$, $N(1710)$, and $N_x(2065)$
already included. The log likelihood values get better by 28, 137,
69, and 73, respectively, which indicates the $N(2080)$ is the
most significant, all the significances are larger than $5
\sigma$. Second, we include $N_x(2065)$ and $N(2080)$ in the high
mass region and add the other three states $N(2000)$, $N(2090)$,
and $N(2100)$ one at a time. The significances of the $N(2100)$
($\Delta S = 38$) and $N(2090)$ ($\Delta S = 30$) are much larger
than $5\sigma$, while $N(2000)$ is $3.9\sigma$ ($\Delta S=14$).
%Including the $N(2100)$ is found to have the best log likelihood
%value, and $N(2090)$ has only a significance of 2.7$\sigma$.
Third, we include $N_x(2065)$, $N(2080)$, and $N(2100)$ in the high mass
region and test whether $N(2000)$ and $N(2090)$ are needed again. The
significances are larger than $5\sigma$ ($\Delta S = 23$) and
$2.7\sigma$ ($\Delta S = 7$), respectively when
$N(2000)$ and $N(2090)$ are included.

Due to the complexity of the high mass $N^*$ states and the limitation
of our data, we are not able to draw firm conclusions on the high
mass region. In the final fit, we include $N_x(2065)$,
$N(2080)$, and $N(2100)$ and take the
differences of with and without $N(2000)$ and $N(2090)$ as
systematic errors.

\subsubsection{The best results up to now}

We summarize the results we have so far: \\

(1) For the three resonances in the $M=1.7$ GeV/c$^2$ mass region
($N(1700)$, $N(1710)$, and $N(1720)$), only $N(1710)$ is significant. \\

(2) The $N_{x}(2065)$ is definitely needed in all cases, and its
    spin-parity favors $\frac{3}{2}^+$. \\

(3) $N_{x}(1885)$ is not significant and therefore is not included in
    the final fit.  \\

(4) For other resonances in the high mass region, $N(2080)$ and
$N(2100)$ are both needed in all cases tried, but the other two
states $N(2000)$ and $N(2090)$ are not very significant and so are
not included in the final fit. \\
%However, the differences with
%and without $N(2000)$ and $N(2090)$ are taken as systematic
%errors.\\

Therefore, we consider $N(940)$, $N(1440)$, $N(1520)$, $N(1535)$,
$N(1650)$, $N(1675)$, $N(1680)$, $N(1710)$, $N_{x}(2065)$, $N(2080)$,
and $N(2100)$ in the fit.\\

Table~\ref{tab:mwopt} lists the optimized masses and widths for some
$N^*$ resonances; the others are fixed to those from
PDG08. Here, only statistical errors are indicated. The fractions of
these states are also listed.

\begin{figure*}
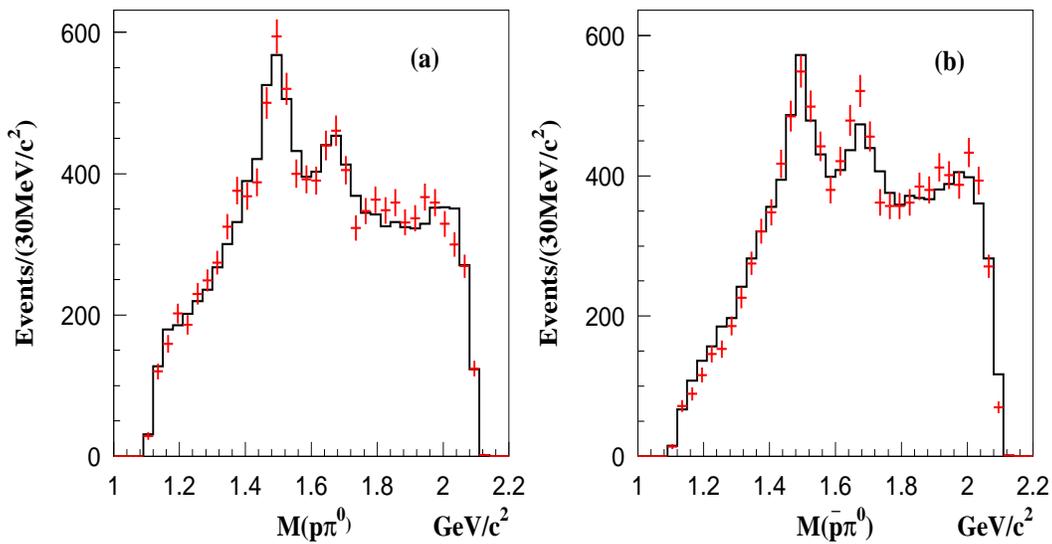

\includegraphics[height=2.8in,width=2.7in]{figu/fit8a.epsi}
\includegraphics[height=2.8in,width=2.7in]{figu/fit8b.epsi}
\caption{The $p \pi^0$ and $\bar p \pi^0$ invariant mass spectra
after optimization of masses and widths. Plot (a) is
$M_{p\pi^0}$, and plot (b) is $M_{\bar p \pi^0}$, where the crosses are
data and histograms are fit results.} \label{fig:afit12}
\end{figure*}

\begin{figure*}
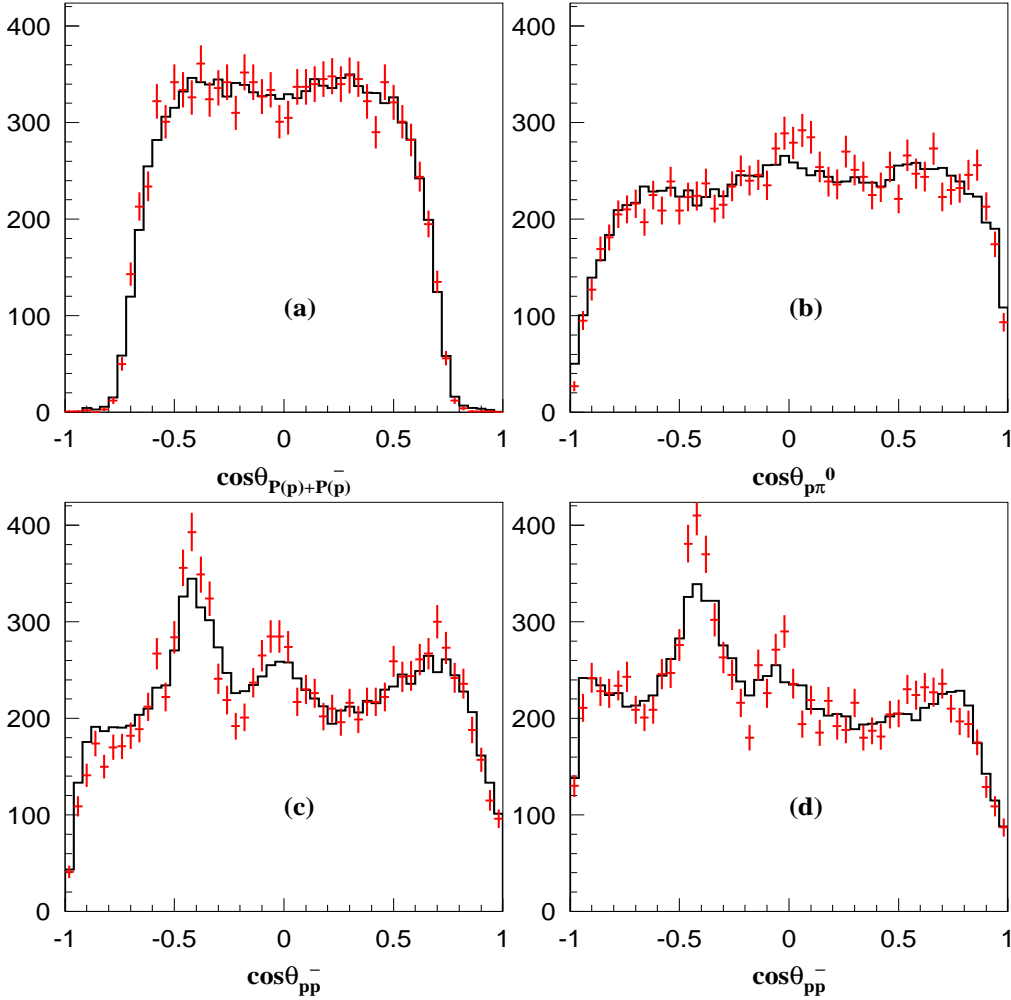

\includegraphics[height=2.6in,width=2.6in]{figu/fit-ang9a.epsi}
\includegraphics[height=2.6in,width=2.6in]{figu/fit-ang9b.epsi}
\includegraphics[height=2.6in,width=2.6in]{figu/fit-ang9c.epsi}
\includegraphics[height=2.6in,width=2.6in]{figu/fit-ang9d.epsi}
\caption{Distributions of (a) the cosine of the sum of the $p$ and
  $\bar{p}$ momenta, (b) cosine of the momentum of the ${p \pi^0}$
  system in the  $p \bar p$ CMS, (c) cosine of the momentum of the $p
    \bar p$ system in the $p \pi^0$ CMS, and (d) cosine of the
  momentum of $p \bar p$ in the $\bar p \pi^0$ CMS. The crosses
  are data and histograms are the fit results.}
\label{fig:afit4}
\end{figure*}

The $M_{p\pi^0}$ and $M_{\bar p \pi^0}$ invariant mass spectra and the
angular distributions after the optimization are shown in
Figs.~\ref{fig:afit12}~(a) and (b) and Fig.~\ref{fig:afit4},
respectively. In Fig.~\ref{fig:afit12} and \ref{fig:afit4}, the
crosses are data and the histograms are the PWA fit projections. The
PWA fit reasonably describes the data.

\begin{table*}[htpb]
\caption{\label{tab:mwopt}Optimized masses and widths, as well as
fractions. Errors shown are statistical only.}
\begin{ruledtabular}
\begin{tabular}{ccccc}
Resonance & Mass(MeV/c$^2$) & Width(MeV/c$^2$) & $J^P$ &Fraction ($\%$)\\
\hline
$N(1440)$ & $1455_{-7}^{+2}$&$316_{-6}^{+5}$&$\frac{1}{2}^+$&16.37\\
 $N(1520)$ & $1513_{-4}^{+3}$ &$127_{-8}^{+7}$&$\frac{3}{2}^-$&7.96\\
 $N(1535)$ & $1537_{-6}^{+2}$ & $135_{-8}^{+8}$&$\frac{1}{2}^-$&7.58\\
 $N(1650)$ & $1650_{-6}^{+3}$&$145_{-10}^{+5}$&$\frac{1}{2}^-$&9.06\\
 $N(1710)$ & $1715_{-2}^{+2}$&$95_{-1}^{+2}$&$\frac{1}{2}^+$&25.33\\
 $N_{x}(2065)$ & $2040_{-4}^{+3}$&$230_{-8}^{+8}$&$\frac{3}{2}^+$&23.39\\
\end{tabular}
\end{ruledtabular}

\end{table*}

\subsubsection{ $N_{x}(1885)$ significance with optimized $N^*$ states}

In the analysis above, the $N_{x}(1885)$ was not found to be
significant. Here its significance is redetermined using the optimized
masses and widths for the $N^*$'s, and it is still only 1.2$\sigma$
($\Delta S = 2.7$).
Therefore, we have the same conclusion: the $N_{x}(1885)$ is not
needed.

\subsubsection{ $N(1900)$}

In PDG08~\cite{pdg}, there is an $N(1900)$($\frac{3}{2}^+$) state
near $N_{x}(2065)$ \cite{pdg}. Our previous results show that if
there is only one $\frac{3}{2}^+$ state in this region, the mass and
width are optimized at $M=2057_{-6}^{+4}$ MeV/c$^2$ and
$\Gamma=220_{-12}^{+11}$ MeV/c$^2$, which are
consistent with those of $N_{x}(2065)$. If $N(1900)$ is
also included in this analysis, {\it i.e.} there are two
$\frac{3}{2}^+$ states in this region, we find that the second
$\frac{3}{2}^+$ state also has a statistical significance much
larger than 5$\sigma$ ($\Delta S = 49$). However, the interference between $N(1900)$
and $N_{x}(2065)$ is about 80\%.  This analysis does not exclude
the possibility that there are two $\frac{3}{2}^+$ states in this region.

\subsubsection{ Search for additional $N^*$ and $\Delta^*$ resonances}

Besides the contributions from the well-established $N^*$ resonances,
there could be smaller contributions from other $N^*$ resonances and
even $\Delta^*$ resonances from isospin violating virtual photon
production.

What might be expected for the isospin violating decay? For the
$J/\psi \rightarrow p \bar p$ decay, the isospin violating fraction
can be estimated using the PDG $J/\psi$ leptonic branching fraction
and the proton electromagnetic form factor $F_{p}(q^2)$\cite{baub} to be
$B(J/\psi \rightarrow \gamma^* \rightarrow p \bar p)$ = $B(J/\psi
\rightarrow l^+ l^-) \times (F_{p}(M_{J/\psi}^2)^2$ = $2.4 \times
10^{-5}$. The total $J/\psi \rightarrow p \bar p$ branching fraction is
$2.2 \times 10^{-3}$\cite{pdg}. This means, the fraction of decays
through a virtual photon in the $J/\psi \rightarrow \gamma^*
\rightarrow p \bar p$ decay mode is close to 1.1\%. For the
non-strange channel, the ratio of photon couplings to isospin 1 and
isospin 0 is 9:1, so the isospin violating part is about 1\% for this
channel.  For the $J/\psi \rightarrow p \bar p \pi^0$ decay, one would
expect a similar isospin violating fraction.

If we add an extra state with different possible spin-parities
($J^P=\frac{1}{2}^{\pm}, \frac{3}{2}^{\pm}, \frac{5}{2}^{\pm}$) in the
large mass (1.65 GeV/c$^2$ to 1.95 GeV/c$^2$) region with widths from
0.05 GeV/c$^2$ to 0.20 GeV/c$^2$ and re-optimize, we find that no
additional $N^*$'s or $\Delta^*$'s with the statistical significance
of greater than 5$\sigma$ are required.

\subsubsection{Search for $\rho(2150)$}

\begin{figure*}
\includegraphics[height=2.6in,width=2.6in]{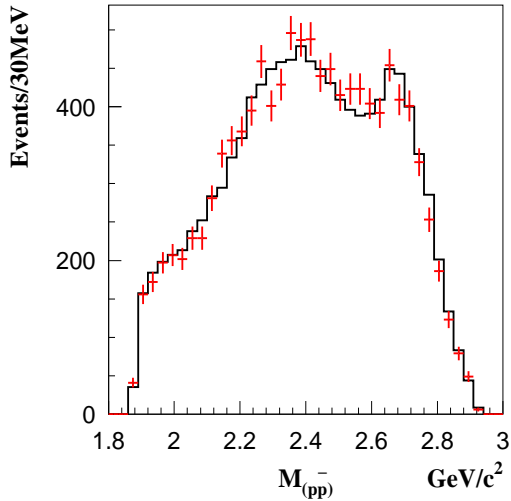}

\caption{The $p \bar p$ invariant mass spectra. Crosses are data and
  histogram is the PWA fit after the optimization of masses and
  widths.} \label{fig:afit3c}
\end{figure*}

A resonance with mass 2149 MeV/c$^2$ and $J^P=1^-$ is listed in
PDG08~\cite{pdg} with the decay $\rho(2150) \rightarrow p \bar p$.
Here, we test whether there is evidence for this decay in our
sample. The significance of this resonance is less than 3$\sigma$
when we vary the width of this state in the fit from 200 to 660
MeV/c$^2$. Therefore, our data do not require this state.
Figure~\ref{fig:afit3c} shows the $p \bar{p}$ invariant mass
spectrum, and there is no clear structure near 2149 MeV/c$^2$.
%
%\begin{table}[htpb]
%\caption{\label{tab:ppb}The significance of possible $\rho(2150)
%\rightarrow p \bar p$ structure versus width.}
%\begin{ruledtabular}
%\begin{tabular}{ccc}
%Width(MeV/c$^2$)&$\Delta S$
%&significance\\
%\hline
 %660        &     -6      &      2.38$\sigma$\\
 %610        &     -6      &      2.22$\sigma$\\
 %560        &     -6      &      2.38$\sigma$\\
 %510        &     -6      &      2.28$\sigma$\\
 %460        &     -6      &      2.25$\sigma$\\
 %410        &     -5      &      2.05$\sigma$\\
 %363        &     -5      &      2.03$\sigma$\\
 %300        &     -4      &      1.82$\sigma$\\
 %250        &     -4      &      1.57$\sigma$\\
% 200        &     -3      &      1.39$\sigma$\\
%\end{tabular}
%\end{ruledtabular}
%\end{table}

\section{Branching fraction of $J/\psi \to p \bar p \pi^0$}

\begin{figure*}
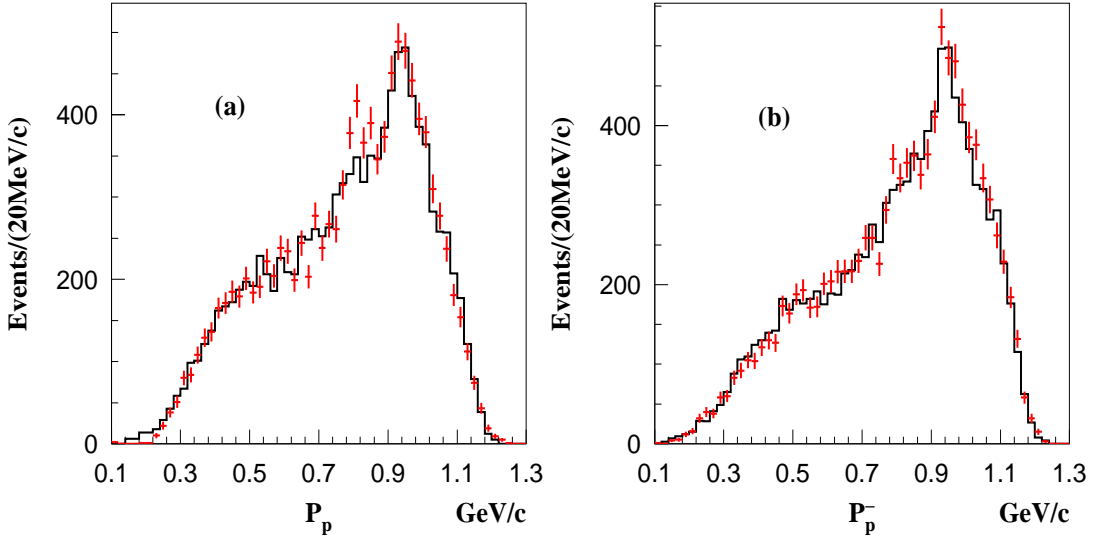

\includegraphics[height=2.8in,width=2.8in]{figu/fdc-simpp11a.epsi}
\includegraphics[height=2.8in,width=2.8in]{figu/fdc-simpp11b.epsi}
\caption{Momentum distributions of $p$ and $\bar p$ in $J/\psi
  \rightarrow p \bar p \pi^0$, where histograms are $J/\psi
  \rightarrow p \bar p \pi^0$ MC simulation using the spin parities
  and fractions of $N^*$ states obtained from our partial wave analysis and
  crosses are data.} \label{fig:fdc-sim}
\end{figure*}

The branching fraction of $J/\psi \to p \bar p \pi^0$ is obtained by
fitting the $\pi^0$ signal (see Fig. 3) with a $\pi^0$ shape obtained
from $J/\psi \rightarrow p \bar p \pi^0$ MC simulation and a
polynomial background. The numbers of fitted signal and background
events are 11,166 and 691, respectively. The efficiency of $J/\psi \to
p \bar p \pi^0$ is determined to be 13.77\% by MC simulation with
all intermediate $N^*$ states being included.
Figures~\ref{fig:fdc-sim}~(a) and (b) show the $p$ and $\bar p$ momentum
distributions, where the histograms are MC simulation of $J/\psi
\rightarrow p \bar p \pi^0$ using the $J^P$'s and fractions of $N^*$
states obtained from our partial wave analysis, and the crosses are data. There
is good agreement between data and MC simulation.

The branching fraction is determined to be:
\begin{equation}
Br(J/\psi \rightarrow p \bar p \pi^0)=(1.33 \pm 0.02~(stat.)) \times
10^{-3},
\end{equation}
which is higher than that in PDG08 \cite{pdg} ($(1.09 \pm 0.09)
\times 10^{-3}$).

\section{Systematic errors}

The systematic errors for the masses and widths of $N^*$ states
mainly originate from the difference between data and MC simulation,
the influence of the interference between $N(2100)$ and other
states, uncertainty of the background, the form-factors, and the
influence of high mass states, as well
as the differences when small components are included or not.

(1) Two different MDC wire resolution simulation models are used
to estimate the systematic error from the data/MC difference.

(2) In this analysis, the interference between
$N(2100)$($\frac{1}{2}^+$) and the low mass regions states such as
$N(940)$($\frac{1}{2}^+$) and $N(1440)$($\frac{1}{2}^+$) can be very
large, even larger than 50\%. We fix the fraction of $N(2100)$ to be
less than 10\% to reduce the interference and then compare its impact
on other resonances. The biggest differences for the masses, widths,
and fractions of the other resonances between fixing the fraction of
$N(2100)$ and floating its fraction are considered
as systematic errors.

(3) Two kinds of backgrounds are considered in the partial wave
analysis, $\pi^0$ sideband and non-interfering phase space.  We
increase the number of background events by 10\%, and take the
changes of the optimized masses and widths as systematic errors.

(4) Equations (12) and (13) are the form factors used in this analysis,
where $\Lambda$ is 2.0 for $N^*$ states with $J^P= \frac{1}{2}$ and
$\Lambda $ is 1.2 for those with $J^P= \frac{3}{2}$ and
$\frac{5}{2}$. Other form factors have also been tried, however their
log likelihood values are much worse than those from the form factors
used here. We also vary the $\Lambda$ values from 2.0 and 1.2 to
1.5. The biggest differences are taken as the form factor systematic
errors.

(5) The effect of using different combinations of states in the high
mass region on the masses and widths of other resonances was
investigated above (see Table~\ref{tab:case}), and the differences
also taken as systematic errors.

Table \ref{tab:sumerr} shows the summary of the systematic errors for
the masses and widths, and the total systematic errors are the sum of
each source added in quadrature.

%During the previous systematic errors analysis, we
%found that the differences of mass, width between some cases and the
%best result are 0 even they are optimized by the step of 1 MeV/c$^2$.
%However it's known to us all the error couldn't be 0, so we enlarge
%the value to be 1 when it's 0.
\begin{table*}
\begin{ruledtabular}
\caption{\label{tab:sumerr} Summary of the systematic errors for
masses and widths of $N^*$ resonances (MeV/c$^2$).}
\begin{tabular}{lcccccccccccc}
Systematic error& \multicolumn{2}{c}{$N(1440)$}&
\multicolumn{2}{c}{$N(1520)$}& \multicolumn{2}{c}{$N(1535)$}&
\multicolumn{2}{c}{$N(1650)$}& \multicolumn{2}{c}{$N(1710)$}&
\multicolumn{2}{c}{$N_{x}(2065)$} \\
&$\Delta M$ & $\Delta \Gamma$&
$\Delta M$  & $\Delta \Gamma$&
$\Delta M$  & $\Delta \Gamma$&
$\Delta M$  & $\Delta \Gamma$&
$\Delta M$  & $\Delta \Gamma$&
$\Delta M$  & $\Delta \Gamma$\\
\hline Data/MC comparison    &3 &14 &2  &13 &2  &11 &4  &12 &1&12 &1
 &19\\
Interference     &12 &25 &2  &23 &3  &22 &25 &15 &15 &2  &10 &20\\
Background uncertainty   &18 &51 &11 &23 &6  &28 &2  &8  &4  &10 &5  &15\\
Different form-factors     &12 &25 &2  &5  &8  &1 &3  &5  &15 &22 &20 &14\\
Different combinations in high mass region     &35 &21 &7  &12 &5  &10 &5  &23 &20 &35 &10 &39\\
Total                  &43 &67 &13 &37 &12 &39 &26 &31 &29 &44 &25 &52\\
\end{tabular}
\end{ruledtabular}
%\footnotetext[1]{Enlarge the value to be 1 since it's 0 in this study}
\end{table*}

\begin{table}
\caption{\label{tab:brerr}
Systematic error for the branching fraction $B(J/\psi \to \pi^0 p \bar{p})$ from different
sources.}
\begin{ruledtabular}
\begin{tabular}{lc}
Sys. error sources & Systematic error(\%)\\
\hline
Wire resolution & 2.18\\
Photon efficiency & 4.00\\
Particle ID & 4.00\\
Mass spectrum fitting & 1.93\\
%$B(\pi^0 \rightarrow \gamma \gamma)$ & 0.04\\
Number of $J/\psi$ events & 4.72\\
Total&7.93\\
\end{tabular}
\end{ruledtabular}
\end{table}

 \begin{table*}[htpb]
\begin{ruledtabular}
\caption{\label{tab:optr}Summary of $N^*$ states optimized results.}
\begin{tabular}{cccccc}
Resonance &Mass(MeV/c$^2$)&width(MeV/c$^2$)& $J^P$ &Fraction(\%)&Branching
fraction ($\times 10^{-4}$)\\
\hline
 $N(1440)$ & $1455_{-7}^{+2}\pm 43$&$316_{-6}^{+5}\pm
67$&$\frac{1}{2}^+$&9.74$\sim$25.93&1.33$\sim$3.54\\
 $N(1520)$ & $1513_{-4}^{+3}\pm 13$ &$127_{-8}^{+7}\pm
37$&$\frac{3}{2}^-$&2.38$\sim$10.92&0.34$\sim$1.54\\
 $N(1535)$ & $1537_{-6}^{+2}\pm 12$ & $135_{-8}^{+8}\pm
39$&$\frac{1}{2}^-$&6.83$\sim$15.58&0.92$\sim$2.10\\
 $N(1650)$ & $1650_{-6}^{+3}\pm 26$&$145_{-10}^{+5}\pm
31$&$\frac{1}{2}^-$&6.89$\sim$27.94&0.91$\sim$3.71\\
 $N(1710)$ & $1715_{-2}^{+2}\pm 29$&$95_{-1}^{+2}\pm
44$&$\frac{1}{2}^+$&4.17$\sim$30.10&0.54$\sim$3.86\\
 $N_{x}(2065)$ & $2040_{-4}^{+3}\pm 25$&$230_{-8}^{+8}\pm
52$&$\frac{3}{2}^+$&7.11$\sim$24.29&0.91$\sim$3.11\\
\end{tabular}
\end{ruledtabular}
\end{table*}

The systematic errors for the branching fraction $B(J/\psi \to \pi^0 p
\bar{p})$ mainly originate from the data/MC discrepancy for the tracking
efficiency, photon efficiency, particle ID efficiency, fitting region
used, the background uncertainty, and the uncertainty in the number of
$J/\psi$ events.

(1) The systematic error from MDC tracking and the kinematic fit,
2.18\%, is estimated by using different MDC wire resolution
simulation models.

(2) The photon detection efficiency has been studied using $J/\psi
\rightarrow \rho \pi$~\cite{smli}. The efficiency difference between
data and MC simulation is about 2\% for each photon. So 4\% is taken
as the systematic error for two photons in this decay.

(3) A clean $J/\psi \rightarrow p \bar p \pi^+ \pi^-$ sample is used
to study the error from proton identification. The error from
the proton PID is about 2\%. So the
total error from PID is taken as 4\% in this decay.

(4) %The changes of the fitting parameters and mass region may lead
%to the change of result.
%In this analysis, these two kinds errors
%are calculated to be the systematic errors from uncertainty of
%background. The signal events is obtained from fitting $\pi^0$ mass
%spectrum. In order to estimate the uncertainty caused by the fitting
%range,
The $\pi^0$ fitting range is changed from 0.04 - 0.3 GeV/c$^2$ to 0.04 - 0.33
GeV/c$^2$, and the difference , 1.28\%, is taken to be the systematic
error from the fitting range. To estimate the uncertainty from the
background shape, we change the background shape from 3rd order
polynomial to other functions. The biggest change, 1.44\%, is taken as
the systematic error.

%(5) The branching fraction of $\pi^0 \rightarrow \gamma \gamma$ is
%$(98.798 \pm 0.032)$\% \cite{pdg}. The systematic error from
%$Br(\pi^0 \rightarrow \gamma \gamma)$ is about 0.04\%.

(5) The total number of $J/\psi$ events determined from inclusive
4-prong hadrons is $(57.70 \pm 2.72) \times 10^6$ \cite{fangss}. The
uncertainty is 4.72\%.

Table~\ref{tab:brerr} lists the different sources of systematic
errors for the branching fraction of $J/\psi \rightarrow p \bar p
\pi^0$. The total systematic error is the sum of each error added
in quadrature.
%So the branching fraction of $J/\psi
%\rightarrow p \bar p \pi^0$ decay mode in our analysis is
%\begin{equation}
%Br(J/\psi \rightarrow p \bar p \pi^0)=(1.33 \pm 0.02 \pm 0.11)
%\times 10^{-3}
%\end{equation}

\section{Summary}

Based on 11,166 $J/\psi \rightarrow p \bar p \pi^0$ candidates
from $5.8 \times 10^7$ BESII $J/\psi$ events, a partial wave amplitude
analysis is performed. A long-sought `missing' $N^*$, which was
observed first by BESII in $J/\psi \to p \bar n \pi^- + c.c.$, is also
observed in this decay with mass and width of $2040_{-4}^{+3}\pm
25$ MeV/c$^2$ and $230_{-8}^{+8}\pm 52$ MeV/c$^2$, respectively. The
mass and width obtained here are consistent with those from $J/\psi
\to p \bar n \pi^- + c.c.$ within errors. Its spin-parity favors
$\frac{3}{2}^+$. The masses and widths of other $N^*$ resonances in
the low mass region are also obtained and listed in
Table~\ref{tab:optr}, where the first errors are statistical and the
second are systematic. The ranges for the fractions of $N^*$ states,
and thus the branching fractions, are given too. From this analysis, we
find that the fractions of each $N^*$ state depend largely on the
$N^*$'s used in the high mass region, the form factors, and Breit-Wigner
parameterizations, as well as the background. We also determine the
$J/\psi \to p \bar p \pi^0$ branching fraction to be $Br(J/\psi
\rightarrow p \bar p \pi^0)=(1.33 \pm 0.02 \pm 0.11) \times 10^{-3}$,
where the efficiency used includes the intermediate $N^*$ and $\bar
N^*$ states obtained in our partial wave analysis.

\section{Acknowledgments}

The BES collaboration thanks the staff of BEPC and computing center
for their hard efforts. This work is supported in part by the
National Natural Science Foundation of China under contracts Nos.
10491300, 10225524, 10225525, 10425523, 10625524, 10521003,
10821063, 10825524, the Chinese Academy of Sciences under contract
No. KJ 95T-03, the 100 Talents Program of CAS under Contract Nos.
U-11, U-24, U-25, and the Knowledge Innovation Project of CAS under
Contract Nos. U-602, U-34 (IHEP), the National Natural Science
Foundation of China under Contract Nos. 10775077, 10225522 (Tsinghua
University), and the Department of Energy under Contract No.
DE-FG02-04ER41291 (U. Hawaii).


\begin{thebibliography}{99}
\bibitem{isgur} N. Isgur, nucl-th/0007008 (2000).
\bibitem{bnl} E. G. Cazzoli, Phys. Rev. Lett. 34, 1125 (1975).
\bibitem{cb1}
S.~E.~Csorna et al., Phys. Rev. Lett., 86, 4243 (2001).
\bibitem{cb2}
B.~Aubert (BABAR Colaboration), Phys. Rev. D 72, 052006 (2005).\\
B.~Aubert (BABAR Colaboration), hep-ex/0607042.\\
B.~Aubert (BABAR Colaboration), Phys. Rev. Lett., 97, 232001 (2006).\\
B.~Aubert (BABAR Colaboration), Phys. Rev. D 74, 011103 (2006).\\
B.~Aubert (BABAR Colaboration), hep-ex/0607086.\\
B.~Aubert (BABAR Colaboration), Phys. Rev. Lett., 98, 012001 (2007).\\
B.~Aubert (BABAR Colaboration), Phys. Rev. D 77, 012002 (2009).\\
\bibitem{cb3}
K.~Abe (BELLE Collaboration), Phys. Rev. Lett., 98, 262001 (2007).\\
R.~Chistov (BELLE Collaboration), Phys. Rev. Lett., 97, 162001 (2006).\\
R.~Mizuk (BELLE Collaboration), Phys. Rev. Lett., 94, 122002 (2005).
\bibitem{cb4}
M.~Iori et al., hep-ex/0701021.
\bibitem{cb5}
C.~Amsler et al., Phys. Lett. B 667, 1 (2008).
\bibitem{bb1}
G.~Bari, et al., Nuovo Cim., A 104, 1787 (1991).\\
G.~Bari, et al., Nuovo Cim., A 104, 571 (1991).
\bibitem{bb2}
T.~Aaltonen (CDF Collaboration), Phys. Rev. Lett., 99, 202001 (2007).\\
T.~Aaltonen (CDF Collaboration), Phys. Rev. Lett., 99, 052002 (2007).
\bibitem{bb3}
V.~M.~Abazov (D0 Collaboration), Phys. Rev. Lett., 99, 052001 (2007).\\
V.~M.~Abazov (D0 Collaboration), arXiv:0808.4142.
\bibitem{baub} B. Aubert et al., Phys. Rev. D73, 012005 (2006).
\bibitem{pdg}C. Amsler et al., Physics Lett. B667, 1 (2008).
\bibitem{scaw} S. Capstick and W. Roberts, Prog. Part. Nucl. Phys. 45, S241 (2000).
\bibitem{nig} N. Isgur and G. Karl, Phys. Rev. D19, 2653 (1979).
\bibitem{scaw2} S. Capstick and W. Roberts, Phys. Rev. D47, 1994 (1993).
\bibitem{plb75} J. Z. Bai et al. (BES
Collaboration), Phys. Lett. B510, 75 (2001).
\bibitem{xbji} M. Ablikim et al. (BES Collaboration), Phys. Rev. Lett. 97, 062001 (2006).
\bibitem{bes2} J. Z. Bai et. al, (BES Collab.), Nucl. Inst. and Meths. A458,
627 (2001).
% BES Collaboration, M.~Ablikim {\em et al.},
%               physics/0503001.
\bibitem{simbes}M. Ablikim et al. (BES Collaboration), Nucl. Instrum. Meth. A552, 344 (2005).
% \bibitem{jpsan} Chen J.C. et al., Phys. Rev. D62, 034003(2000),
%\bibitem{mbnc} M. Benmerrouche, N.C.Mukhopadhyay and J.F.Zhang, Phys. Rev. Lett.
%77, 4716 (1996); Phys. Rev. D51, 3237 (1995),
%\bibitem{mgoet} M.G.Olsson and E.T.Osypowski, Nucl. Phys. B87, 399 (1975); Phys.
%Rev. D17, 174 (1978); M.G.Olsson et al., ibid. 17,2938 (1978),
\bibitem{wrjs} W. Rarita and J. Schwinger, Phys. Rev. 60, 61 (1941).
\bibitem{whl} W. H. Liang, P. N. Shen, J. X. Wang and B. S. Zou, J. Phys. G28, 333 (2002).
\bibitem{suc3}S. U. Chung, Phys. Rev. D48, 1225 (1993).
\bibitem{mbnc} M. Benmerrouche, N. C. Mukhopadhyay and J. F. Zhang, Phys. Rev. Lett.
77, 4716 (1996); Phys. Rev. D51, 3237 (1995).
\bibitem{mgoet} M. G. Olsson and E. T. Osypowski, Nucl. Phys. B87, 399 (1975); Phys.
Rev. D17, 174 (1978); M. G. Olsson, E. T. Osypowski and E. H.
Monsay, Phys. Rev D17,2938 (1978).
\bibitem{cfncs} C. Fronsdal, Nuovo Cimento Sppl. 9, 416 (1958); R. E.
Behrends and C. Fronsdal, Phys. Rev. 106, 345 (1957).
\bibitem{suc2} S. U. Chung, Spin Formalisms, CERN Yellow
Report 71-8 (1971); S. U. Chung, Phys. Rev. D48, 1225 (1993); J.
J. Zhu and T. N. Ruan, Communi. Theor. Phys. 32, 293, 435 (1999).

%\bibitem{jxwang} J.X.Wang, Comput. Phys. Commun. 77, 263 (1993),

\bibitem{larfd} L. Adler and
R. F. Dashen, Current Algebra and Application to Particle
Physics (Benjamin, New York, 1968); B. W. Lee, Chiral Dynamics (Gordon
and Breach, New York, 1972).

\bibitem{tpv181} T. P. Vrana, S. A. Dytman and T. S. H. Lee, Phys. Rept. 328,
181 (2000).

\bibitem{liangwh} Liang Wei-hong. Ph.D thesis, Institute of High Energy Physics,
Chinese Academy of Science, 2002 (in Chinese); G.Penner and U. Mosel, Phys. Rev. C66, 055211 (2002); W. H. Liang et al., Eur. Phys. J. A21, 487 (2004).
\bibitem{smli} S. M. Li et al., HEP $\&$ NP 28, 859 (2004) (in Chinese).
\bibitem{fangss} Fang S.S. et al., HEP $\&$ NP 27, 277 (2003) (in Chinese).
\end{thebibliography}
\end{document}